\documentclass[conference]{IEEEtran}
\IEEEoverridecommandlockouts
\usepackage{cite}
\usepackage{amsmath,amssymb,amsfonts}
\usepackage{algorithmic}
\usepackage{graphicx}
\usepackage{textcomp}
\usepackage{ifpdf}
\usepackage{enumitem}
\usepackage{varwidth}
\usepackage{amsmath}
\usepackage[table]{xcolor}
\usepackage[framemethod=tikz]{mdframed} 
\usepackage{enumitem}
\usepackage{microtype}
\def\BibTeX{{\rm B\kern-.05em{\sc i\kern-.025em b}\kern-.08em
    T\kern-.1667em\lower.7ex\hbox{E}\kern-.125emX}}

\usepackage{tabularx}
\usepackage{graphicx}
\usepackage{textcomp}
\usepackage[ampersand]{easylist}
\usepackage{multicol}
\usepackage{multirow}

\usepackage{threeparttable}
\usepackage{tabularx}

\usepackage{rotating}
\usepackage[roman]{parnotes}
\usepackage[many]{tcolorbox}
\usepackage{soul}
\usepackage{xcolor}
\usepackage{comment}
\usepackage{booktabs}
\usepackage{colortbl}
\usepackage{verbatim}
\usepackage{xspace}
\usepackage{xcolor}
\xspaceaddexceptions{]\}}

\usepackage{hyperref}
\usepackage{footmisc}
\usepackage{array}
\usepackage[export]{adjustbox}
\usepackage{longfbox}
\usepackage{commath}
\usepackage[normalem]{ulem}
\usepackage{mdframed,lipsum}
\usepackage{cleveref}
\usepackage{makecell}

\usepackage{fontawesome}

\usetikzlibrary{calc}

\definecolor{labelColor}{RGB}{56,152,236}

\newcommand{\greenup}{\color[HTML]{228B22}{\faArrowUp}}
\newcommand{\reddown}{\color[HTML]{C70039}{\faArrowDown}}

\definecolor{labelColorBig}{RGB}{90, 79, 207}

\usepackage{tikz}

\usepackage[framemethod=tikz]{mdframed} %

\def\BibTeX{{\rm B\kern-.05em{\sc i\kern-.025em b}\kern-.08em
    T\kern-.1667em\lower.7ex\hbox{E}\kern-.125emX}}
\begin{document}


\title{The State of Diversity and Inclusion in Apache: A Pulse Check \\
\thanks{We thank all survey respondents for contributing to this research. This work is supported by the National Science Foundation under Grant No.:NSF1901031, NSF1900903, NSF2236198, NSF2235601 and Google award for inclusion research Program.}
}

\makeatletter
\newcommand{\linebreakand}{%
  \end{@IEEEauthorhalign}
  \hfill\mbox{}\par
  \mbox{}\hfill\begin{@IEEEauthorhalign}
}
\makeatother

\author{\IEEEauthorblockN{Zixuan Feng}
\IEEEauthorblockA{\textit{School of Electrical Engineering and Computer Science} \\
\textit{Oregon State University}\\
Corvallis, Oregon, USA \\
fengzi@oregonstate.edu}
\and
\IEEEauthorblockN{Mariam Guizani}
\IEEEauthorblockA{\textit{School of Electrical Engineering and Computer Science} \\
\textit{Oregon State University}\\
Corvallis, Oregon, USA \\
guizanim@oregonstate.edu}%
\linebreakand 
\IEEEauthorblockN{Marco A. Gerosa}
\IEEEauthorblockA{\textit{School of Informatics, Computing, and Cyber Systems} \\
\textit{Northern Arizona University}\\
Flagstaff, Arizona, USA \\
marco.gerosa@nau.edu}
\and
\IEEEauthorblockN{Anita Sarma}
\IEEEauthorblockA{\textit{School of Electrical Engineering and Computer Science} \\
\textit{Oregon State University}\\
Corvallis, Oregon, USA \\
anita.sarma@oregonstate.edu}
}

\maketitle

\begin{abstract}
Diversity and inclusion (D\&I) in open source software (OSS) is a multifaceted concept that arises from differences in contributors' gender, seniority, language, region, and other characteristics. D\&I has received growing attention in OSS ecosystems and projects, and various programs have been implemented to foster contributor diversity. However, we do not yet know how the state of D\&I is evolving. By understanding the state of D\&I in OSS projects, the community can develop new and adjust current strategies to foster diversity among contributors and gain insights into the mechanisms and processes that facilitate the development of inclusive communities. In this paper, we report and compare the results of two surveys of Apache Software Foundation (ASF) contributors conducted over two years (n=624 \& n=432), considering a variety of D\&I aspects. We see improvements in engagement among those traditionally underrepresented in OSS, particularly those who are in gender minority or not confident in English. Yet, the gender gap in the number of contributors remains. We expect this study to help communities tailor their efforts in promoting D\&I in OSS.
\end{abstract}

.



\section{Introduction}
\label{sec:intro}

Improving the state of Diversity and Inclusion (D\&I) has become an increasingly important mandate for Open Source Software (OSS) communities \cite{bosu2019diversity, guizani2022perceptions, trinkenreich2022women}. A lack of diversity can result in projects missing out on a broad range of backgrounds, qualifications, and perspectives. Increasing diversity is also a ``social good" mandate, as a lack of diversity means that individuals from underrepresented groups miss out on the learning and potential job opportunities afforded by OSS~\cite{marlow2013impression, singer2013mutual}.

Diversity is a multifaceted concept. Projects can be diverse in terms of demographics such as age, gender, seniority, or ethnicity, as well as participant backgrounds (role, expertise, personality, or cognitive styles) \cite{guizani2022perceptions, ortu2017diverse, prana2021including}. While prior studies have looked at D\&I in OSS, these studies have typically focused on specific aspects of diversity \cite{marlow2013impression, singer2013mutual, prana2021including, trinkenreich2022women, guizani2022perceptions} and provide a single snapshot in time.

OSS communities have started several initiatives to improve the state of D\&I. For example, the Linux \cite{linux} and Apache Software  Foundations (ASF) \cite{EDIgroup} have launched D\&I initiatives that implement mentorship, scholarship, training, and certification programs and promote D\&I best practices such as inclusive naming. 

Thus far, to the best of our knowledge, no research empirically investigates the evolution of the state of D\&I in OSS. Without a pulse check on the state of D\&I, OSS community lacks an understanding of what has improved and what requires more attention. In our work, we investigate the following: 

\noindent\textbf{RQ: How much do contributors from minorities...}
\begin{enumerate}[wide=0pt, start=1,label={\bfseries Sub-RQ\arabic*:}]
\item ...engage in OSS activities?
\item ...face challenges?
\item ...perceive the state of D\&I? 
\end{enumerate}

To answer the research questions, we performed a longitudinal case study within the ASF projects. Thus far, we have conducted two cross-sectional surveys, one in 2020 and the other in 2022. The surveys use the same Likert-scale questions to allow us to evaluate the difference in contributors' perspectives and engagement over time, serving as a pulse check. 

Our work complements prior research by adopting multidimensional attributes to investigate changes in contributors' perceptions of diversity through four perspectives (gender, proficiency in English, seniority, and country of residence), as these four perspectives are significant and have been demonstrated to impact D\&I in OSS.  We believe deploying a pulse check on this multidimensional approach provides a nuanced and comprehensive understanding of the complex factors that lead to D\&I in OSS communities.

Our findings provide an in-depth analysis of the present state of D\&I in OSS communities. Specifically, we offer insights into the following: 1) a pulse check of the state of D\&I engagement across five distinct OSS daily activities; 2) the current state of challenges faced by contributors regarding their D\&I backgrounds; 3) the perception of the contemporary state of D\&I in four areas: role stereotypes, the ability to contribute, being represented, and following a code of conduct.

This pulse check can help OSS communities make informed decisions about proceeding with their D\&I initiatives. Understanding the state of D\&I in OSS can inform practitioners when developing and adjusting programs and initiatives to promote contributor diversity, ultimately benefiting OSS community and software industry by increasing diversity and equity. Additionally, studying the evolution of D\&I in OSS can help researchers in OSS understand inclusive community development. It can assist in uncovering diversity and inclusion factors and inform the design of future studies and experiments, leading to more effective and efficient research in OSS. These insights can also improve future research by informing the design of the study methodology.

\section{Related Work}
\label{sec:relatedwork}

In recent years, diversity in OSS projects has earned widespread attention, mirroring the growing awareness of the importance of diversity in broader workplace settings. Nevertheless, most software engineering research on diversity focuses on gender diversity, mostly on women contributors \cite{bosu2019diversity, ortu2017diverse, trinkenreich2022women, trinkenreich2021please}. By conducting a systematic literature survey, Trinkenreich et al. \cite{trinkenreich2022women} summarized ten strategies to mitigate challenges women contributors  encounter in OSS, such as ``Promote women-specific groups and events'' and ''Encourage and be welcoming to women''. A recent study has created and investigated a systematic debugging process to empower project leaders to reduce gender-biased contribution barriers within their project's workflow \cite{guizani2022debug}. Research has shown that teams with a more diverse gender composition perform better \cite{ortu2017diverse}. Perez \cite{perez2019invisible} discusses data biases that affect women.

Seniority has also been investigated in a few other studies. Senior contributors often lack the social motivation of their younger contributors \cite{morrison2016veteran}, and younger contributors are confronted with a wide range of technical and social issues \cite{davidson2014older}. When it comes to mitigating community smells, many contributors believe that experience is more significant than gender diversity \cite{catolino2019gender}. Steinmacher et al.~\cite{steinmacher2018let} present guidelines for OSS communities and newcomers to help mitigate barriers, and Guizani et al. \cite{guizani2022Dashboard} designed a maintainer dashboard to help attract and retain newcomers. Mentoring is also a strategy often used to support newcomer onboarding \cite{balali2018newcomers}.

Researchers have found that geo-location is an additional factor that impacts the contributors' activities; having a pull request submitted by someone in the same country as the person doing the integrating results in a higher acceptance rate \cite{rastogi2016geographical}. By building a conceptual model of the challenges faced by contributors at large OSS organizations, researchers have reported that some of the social challenges that contributors face are related to geographical isolation and the lack of representation of non-western countries and suggested diversifying event locations \cite{guizani2021long}. Recent research highlighted that gender diversity is low across all parts of the world, and while there is some variation across regions, the difference is insignificant \cite{prana2021including}. A study suggested using several communication methods to address geographic challenges such as cultural differences, time zone issues, and English language proficiency. Encouraging local language and culture groups can help contributors feel at home, especially newcomers \cite{balali2018newcomers}. 

Researchers in OSS communities have put in a lot of effort to make communities more welcoming and accepting of contributors of diverse backgrounds. This paper aims to provide a pulse check of the evolution of contributors' perspectives and engagement over time. We hope this study will help to guide new and existing D\&I initiatives. 

\section{method}

Longitudinal studies focus on discovering trends or changes in the characteristics of the studied population at both the aggregate and individual levels \cite{menard2002longitudinal}. We are conducting a prospective longitudinal study to observe the evolution of D\&I in Apache projects. Thus far, we have conducted two surveys over the course of two years.

We used Apache as our case study as it is the world's largest OSS foundation with over 430k people and more than 350 projects and initiatives~\cite{ASFwebsite}. Apache projects are investing in improving the state of D\&I and collaborated with us in designing the instruments. The Apache D\&I committee helped validate and pilot our survey questions and gave our survey legitimacy for the community. The Apache D\&I committee comprises 18 experienced contributors, including committers, the Project Management Committee (PMC), and board members. 

\subsection{Survey design}

\begin{table*}[tbp]
\centering
\caption{Likert-Scale sub questions}
\resizebox{7in}{!}{
\begin{tabular}{lll}
\rowcolor[HTML]{EFEFEF} 
\textbf{\begin{tabular}[c]{@{}l@{}}Q5. How often do you engage in the \\ following activities in the ASF?\end{tabular}} & \textbf{\begin{tabular}[c]{@{}l@{}}Q6. Do you face challenges when participating  \\ in the ASF (e.g., language differences,technical \\ expertise, cultural differences, etc)?\end{tabular}} & \cellcolor[HTML]{EFEFEF}\textbf{\begin{tabular}[c]{@{}l@{}}Q7. Thinking about your current \\ project(s), please rate the following statements\end{tabular}} \\
Contributing/reviewing code & \textbf{\begin{tabular}[c]{@{}l@{}}*If not ``No challenges'', follow-up question \\ (only in 2022 survey ): How often do you face \\ the following challenges when participating \\ in the ASF?:\end{tabular}} & \begin{tabular}[c]{@{}l@{}}Other members of the project see me as a \\ parental figure\end{tabular} \\
\rowcolor[HTML]{EFEFEF} 
\begin{tabular}[c]{@{}l@{}}Creating or maintaining \\ documentation\end{tabular} & \begin{tabular}[c]{@{}l@{}}Process related   challenges with getting \\ started on the project\end{tabular} & \cellcolor[HTML]{EFEFEF}I am expected to take care of others than is usual \\
\begin{tabular}[c]{@{}l@{}}Participating in decision making \\ about the project and  development\end{tabular} & \begin{tabular}[c]{@{}l@{}}Process related challenges with navigating \\ the contribution process\end{tabular} & \begin{tabular}[c]{@{}l@{}}I have a hard time following discussions because of \\ technical jargon.\end{tabular} \\
\rowcolor[HTML]{EFEFEF} 
Serving as a community organizer & \begin{tabular}[c]{@{}l@{}}Process related challenges with reception\\ issues in the project\end{tabular} & \cellcolor[HTML]{EFEFEF}\begin{tabular}[c]{@{}l@{}}I feel some members of the community are patronizing \\ to me.\end{tabular} \\
Mentoring other contributors & Process related challenges with licenses & I have an equal chance to get contributions accepted \\
\rowcolor[HTML]{EFEFEF} 
 & \begin{tabular}[c]{@{}l@{}}Social Challenges with communication\\ styles\end{tabular} & \cellcolor[HTML]{EFEFEF}Nothing keeps me from contributing to the project \\
 & \begin{tabular}[c]{@{}l@{}}Social Challenges with feeling imposter \\ syndrome/ fear of making mistakes\end{tabular} & I have a solid network of open-source peers \\
\rowcolor[HTML]{EFEFEF} 
 & \begin{tabular}[c]{@{}l@{}}Social Challenges with facing a lack of \\ recognition\end{tabular} & \cellcolor[HTML]{EFEFEF}\begin{tabular}[c]{@{}l@{}}It was easy to find a mentor with whom I felt \\ comfortable\end{tabular} \\
 & \begin{tabular}[c]{@{}l@{}}Social Challenges with toxic/ unwelcoming\\ environment\end{tabular} & The PMC represents a diverse set of people \\
\rowcolor[HTML]{EFEFEF} 
 & \begin{tabular}[c]{@{}l@{}}Social Challenges with located in a different\\ country/from a different nationality\end{tabular} & \cellcolor[HTML]{EFEFEF}I feel represented in the decision making group \\
 & Social Challenges with cultural   differences & \begin{tabular}[c]{@{}l@{}}I felt safer and more empowered to fully participate \\ in this project because it followed the code of conduct\end{tabular} \\
\rowcolor[HTML]{EFEFEF} 
 & Technical related challenges with  documentation & \cellcolor[HTML]{EFEFEF} \\
 & Technical related challenges with technical Hurdles & 
\end{tabular}}
\label{tab:survey}
\end{table*}













Our survey comprises seven questions (see supplementary for the survey questions~\cite{supply}), a mix of multiple-choice and Likert scale questions (Table \ref{tab:survey}). We reused existing questions when possible. 
The four demographic questions (gender identification, seniority in Apache projects, country of residence, and English confidence) were adapted from the 2016 ``Apache Committer Diversity Survey'' \cite{chaossmetrics} and the ``Open Demographics Survey'' \cite{opendemographicsdoc}. The remaining questions are as follows: Q5 aims to understand the activities that contributors frequently engage in to understand the diversity of contributions, Q6 aims to understand the challenges that contributors face when contributing to Apache projects, and Q7 aims to understand contributors' experiences about aspects of D\&I.
In 2022, if respondents did not select ``not a challenge'' for Q6, we added a follow-up question to understand the frequency with which contributors face process, technical, and social challenges when contributing to Apache projects. The follow-up question was adapted from the high-level categories of the conceptual model of challenges that contributors face in a large OSS organization found in prior literature \cite{guizani2021long}. Our goal was to conduct a more in-depth analysis of the difficulties that arise when individuals contribute to OSS projects.
A cross-sectional analysis of the 2020 survey results are reported in \cite{guizani2022perceptions}.

\subsection{Data collection}

We used GPLv2-licensed Lime Survey \cite{Lime} to deploy our survey. With the help of Apache project managers, we sent out invitations to all ``apache.org'' email addresses and shared the survey on Apache developer mailing lists. Respondents were provided with a consent page that explained the purpose of the survey, the data collection and usage process, and provided a point of contact. Each of the two surveys was open for 45 days. We followed Apache's privacy policies and did not collect identifying information or IP addresses.

Based on an estimated total community size of 7500 contributors, we obtained 624 responses in the 2020 survey, corresponding to an 8.32\% response rate, and 432 responses in 2022, corresponding to a 5.76\% response rate (see Table \ref{tab:demo}). These response rates are consistent with other studies in software engineering and OSS \cite{feng2022case, steinmacher2021being}. The reported demographic or survey respondents are consistent across the two surveys ($<5$\% differences for all demographic attributes).

\begin{table}[tbp]
\caption{Demographics of the survey respondents (2020 \& 2022)}
\centering
\resizebox{\columnwidth}{!}{
\begin{tabular}{lllll}
\rowcolor[HTML]{EFEFEF} 
\multicolumn{1}{l|}{\cellcolor[HTML]{EFEFEF}\textbf{Demographics}} & \multicolumn{2}{c|}{\cellcolor[HTML]{EFEFEF}\textbf{2020}} & \multicolumn{2}{c}{\cellcolor[HTML]{EFEFEF}\textbf{2022}} \\ \hline
\multicolumn{5}{c}{\textbf{Gender}} \\ \hline
\rowcolor[HTML]{EFEFEF} 
\multicolumn{1}{l|}{\cellcolor[HTML]{EFEFEF}Men} & 545 & \multicolumn{1}{l|}{\cellcolor[HTML]{EFEFEF}87.34\%} & 374 & 86.57\% \\
\multicolumn{1}{l|}{Women+non-binary+self-describe} & 45 & \multicolumn{1}{l|}{7.21\%} & 29 & 6.71\% \\
\rowcolor[HTML]{EFEFEF} 
\multicolumn{5}{c}{\cellcolor[HTML]{EFEFEF}\textbf{Senority}} \\ \hline
\multicolumn{1}{l|}{less than 1 year (Newcomer)} & 66 & \multicolumn{1}{l|}{10.58\%} & 28 & 6.48\% \\
\rowcolor[HTML]{EFEFEF} 
\multicolumn{1}{l|}{\cellcolor[HTML]{EFEFEF}greater than 1 year} & 553 & \multicolumn{1}{l|}{\cellcolor[HTML]{EFEFEF}88.62\%} & 404 & 93.52\% \\
\multicolumn{5}{c}{\textbf{English Proficiency}} \\ \hline
\multicolumn{1}{l|}{Comfortable using English} & 551 & \multicolumn{1}{l|}{88.30\%} & 364 & 84.26\% \\
\rowcolor[HTML]{EFEFEF} 
\multicolumn{1}{l|}{\cellcolor[HTML]{EFEFEF}Not comfortable using English} & 56 & \multicolumn{1}{l|}{\cellcolor[HTML]{EFEFEF}8.97\%} & 31 & 7.18\% \\
\multicolumn{5}{c}{\textbf{Region}} \\ \hline
\multicolumn{1}{l|}{\begin{tabular}[c]{@{}l@{}}Western countries \\ (N. America + Europe)\end{tabular}} & 508 & \multicolumn{1}{l|}{81.41\%} & 340 & 78.70\% \\
\rowcolor[HTML]{EFEFEF} 
\multicolumn{1}{l|}{\cellcolor[HTML]{EFEFEF}Not western countries} & 93 & \multicolumn{1}{l|}{\cellcolor[HTML]{EFEFEF}14.90\%} & 68 & 15.74\%
\end{tabular}}
\label{tab:demo}
\end{table}

\subsection{Data analysis}
\textit{Demographics grouping:}
The first step in our analysis was to group respondents into the majority and minority groups for each demographic attribute. \Cref{tab:demo} presents the demographic distributions per attribute.

When considering \textit{gender}, the survey asked respondents to select their gender identity (options: man, woman, gender variant/non-conforming/non-binary, prefer to self-describe, and prefer not to say). Respondents who selected ``man" were in the majority; we grouped as ``gender-minority'' all other respondents except those who selected ``prefer not to say". We decided not to include the latter in our grouping, as we are interested in understanding the perspective of underrepresented genders and ``prefer not to say" does not give us the gender. Table \ref{tab:demo} presents the gender distribution of survey respondents in 2020 and 2022. 

The next attribute of interest is \textit{newcomers} to understand their perspectives. To remain innovative it is important for projects to attract new contributors who bring fresh ideas and bring a diversity of thought \cite{steinmacher2013newcomers}. We classified newcomers as those who have ``less than 1-year" experience in Apache projects (see \Cref{tab:demo}).

The default communication language among Apache projects is English, so we analyze the perspective of respondents from \textit{English proficiency} attributes. Non-native English speakers may have difficulty following discussions, especially those involving  idioms and technical jargon \cite{prana2020including}. In the survey, we asked respondents about their English skills from four perspectives, including questions regarding confidence in (1) speaking (face-to-face), participating in (2) technical or (3) non-technical discussion, and (4) conducting code reviews. Each question included Likert-scale options of very confident, confident, average, uncomfortable, and not confident. We averaged the responses across the four sub-questions. Respondents who scored ``average'' or above for all four questions were classified as ``Comfortable using English''; otherwise, they were classified as ``Not comfortable using English'' (see \Cref{tab:demo}).

Another aspect of diversity is regional diversity, which can serve as a proxy for different cultures and communication styles. Respondents were located across 65 countries (2020: 53 countries; 2022: 46 countries), with the majority coming from North America and Europe; the rest were mainly distributed across Asian countries, including India, China, and Japan. Given this distribution, we categorized regions into ``western" and ``not-western" countries. Western country culture is typically more individualistic, whereas Eastern culture is hierarchical and depends on consensus building when making decisions \cite{schwartz2006theory, bae2000organizational}.

\textit{Question response grouping}:
We analyzed the proportion of responses for each of the listed questions (see \Cref{tab:survey}) per demographic attributes. 

To analyze the Likert-scale question Q5, we grouped the responses ``more than once a week'' and ``more than once a month" into \textit{often} and ``less than once a month" and ``never" into \textit{rarely to never}. We compare, between 2020 and 2022, the proportion of each demographic that \textit{often} engages in each of the listed activities (e.g., the proportion of gender minorities answering \textit{often} in the 2020 survey vs. the proportion of gender minorities answering \textit{often} in 2022). 

For Q6 (challenges), we grouped the responses ``many challenges" and ``several challenges'' into \textit{numerous challenges} and ``a few challenges" and ``no challenges'' into \textit{nearly no challenges}. Similarly, for Q7, we grouped the responses ``completely agree" and ``agree" into \textit{agree} and compared the proportion of each demographic that ``agreed" with each of the listed statements. 

\label{sec:method}
\section{Results And Recommendations}

We performed a pulse check on (1) engagement in activities (Q5), (2) challenges in OSS (Q7), and (3) perception of the state of D\&I (Q10) based on our demographic attributes of interest (gender, seniority, English proficiency, and region). In the following, we present the comparisons of the two editions of the survey and provide recommendations for practice. While there is still a significant gap in the representation of minority groups across all demographic attributes (\Cref{tab:demo}), in this paper, we focus on the participation trends within each group to investigate the evolving state of D\&I in OSS. 

\subsection{Engagement in OSS activities}

\begin{table*}[tbp]
\caption{Comparison of proportional activity engagement between 2020 and 2022 disaggregated by diversity lens.}
\centering
\resizebox{7in}{!}{
\begin{tabular}{lrl|rl|rl|rl|rl|rl|rl|rl}
 & \multicolumn{2}{c|}{\textbf{Man}} & \multicolumn{2}{c|}{\textbf{Minority gender}} & \multicolumn{2}{c|}{\textbf{Newcomer}} & \multicolumn{2}{c|}{\textbf{Not newcomer}} & \multicolumn{2}{c|}{\textbf{\begin{tabular}[c]{@{}c@{}}Comfortable using \\ English\end{tabular}}} & \multicolumn{2}{c|}{\textbf{\begin{tabular}[c]{@{}c@{}}Not comfortable using \\ English\end{tabular}}} & \multicolumn{2}{c|}{\textbf{\begin{tabular}[c]{@{}c@{}}From western \\ countries\end{tabular}}} & \multicolumn{2}{c}{\textbf{\begin{tabular}[c]{@{}c@{}}Not from western \\ countries\end{tabular}}} \\
\rowcolor[HTML]{EFEFEF} 
\textbf{\begin{tabular}[c]{@{}l@{}}Contributing/ reviewing \\ code\end{tabular}} & \greenup & {\color[HTML]{228B22} 1.62\%} & \greenup & {\color[HTML]{228B22} 30.52\%} & \greenup & {\color[HTML]{228B22} 19.87\%} & \greenup & {\color[HTML]{228B22} 3.20\%} & \greenup & {\color[HTML]{228B22} 3.97\%} & \greenup & {\color[HTML]{228B22} 3.41\%} & \greenup & {\color[HTML]{228B22} 2.91\%} & \greenup & {\color[HTML]{228B22} 7.37\%} \\
\textbf{\begin{tabular}[c]{@{}l@{}}Creating/maintaining \\ documentation\end{tabular}} & \greenup & {\color[HTML]{228B22} 3.34\%} & \greenup & {\color[HTML]{228B22} 5.95\%} & \reddown & {\color[HTML]{C70039} -4.64\%} & \greenup & {\color[HTML]{228B22} 4.56\%} & \greenup & {\color[HTML]{228B22} 2.24\%} & \greenup & {\color[HTML]{228B22} 11.13\%} & \greenup & {\color[HTML]{228B22} 0.74\%} & \greenup & {\color[HTML]{228B22} 18.44\%} \\
\rowcolor[HTML]{EFEFEF} 
\textbf{\begin{tabular}[c]{@{}l@{}}Participating in decision \\ making about the project \\ development\end{tabular}} & \greenup & {\color[HTML]{228B22} 4.13\%} & \reddown & {\color[HTML]{C70039} -2.41\%} & \greenup & {\color[HTML]{228B22} 7.68\%} & \greenup & {\color[HTML]{228B22} 1.02\%} & \greenup & {\color[HTML]{228B22} 3.68\%} & \greenup & {\color[HTML]{228B22} 1.67\%} & \greenup & {\color[HTML]{228B22} 0.35\%} & \greenup & {\color[HTML]{228B22} 19.85\%} \\
\textbf{\begin{tabular}[c]{@{}l@{}}Serving as a community \\ organizer\end{tabular}} & \greenup & {\color[HTML]{228B22} 2.46\%} & \reddown & {\color[HTML]{C70039} -11.07\%} & \reddown & {\color[HTML]{C70039} -9.34\%} & \greenup & {\color[HTML]{228B22} 0.60\%} & \reddown & {\color[HTML]{C70039} -0.43\%} & \reddown & {\color[HTML]{C70039} -2.23\%} & \reddown & {\color[HTML]{C70039} -3.24\%} & \greenup & {\color[HTML]{228B22} 17.98\%} \\
\rowcolor[HTML]{EFEFEF} 
\textbf{\begin{tabular}[c]{@{}l@{}}Mentoring other \\ contributors\end{tabular}} & \greenup & {\color[HTML]{228B22} 7.54\%} & \greenup & {\color[HTML]{228B22} 4.16\%} & \greenup & {\color[HTML]{228B22} 13.04\%} & \greenup & {\color[HTML]{228B22} 5.75\%} & \greenup & {\color[HTML]{228B22} 5.81\%} & \greenup & {\color[HTML]{228B22} 25.56\%} & \greenup & {\color[HTML]{228B22} 4.02\%} & \greenup & {\color[HTML]{228B22} 23.90\%}
\end{tabular}}
\label{tab:engagement}
\end{table*}

\Cref{tab:engagement} shows the proportion differences between the answers to the two surveys, disaggregated by the diversity attributes. We considered engagement in OSS activities in which respondents responded ``often" (more than once a week or more than once a month). The {\greenup} and {\reddown} indicate, respectively, a percentage increase and decrease between responses in 2020 and 2022. 

We can see an overall positive trend in \Cref{tab:engagement}, with more respondents engaging in the listed activities. The one exception is serving as community organizers, which sees a reduction. The maximum decrease is for those from gender minorities (11.07\%). This means that 11.07\% fewer respondents, who considered themselves as gender minority, participated as community organizers in 2022 compared to 2020. A reduction in this activity also occurs for newcomers and western respondents. This can be an impact of the COVID-19 pandemic with public event cancellations and travel limitations \cite{ford2021tale}, which might have affected the answers to the 2022 survey. However, a higher proportion of respondents from non-western countries served as community organizers (17.98\% increase).

Next, we discuss the engagement of contributors per demographic attributes. 

\textbf{Gender:} A bias against women arises from role incongruity---widespread cultural associations link men, but not women, with the intellectual aptitude required to work on computer science and OSS \cite{singh2019open, leslie2015expectations}. Such bias can result in fewer women making code contributions. It is heartening to see that when considering activities related to contributing or reviewing code, there is a 30.52\% increase in gender minority participation in 2022 compared to 2020 (there is also a slight increase--1.62\%--among men contributors who participate in contributing/reviewing code). 

There is also an increase among all genders in creating or maintaining documentation. This can be an instance of more contributors, perhaps newcomers, who are participating in OSS in a non-code capacity. The importance of non-code contributions, especially by women contributors, has been increasingly getting visibility and recognition \cite{trinkenreich2020hidden}.

Another bias that can impede women contributors is that women are stereotyped as nurturing, caring, and protective. As a result, women contributors are expected to play the primary caregiver role in their communities~\cite{ kaplan1994woman}, which can take away time from making code-related contributions~\cite{balali2018newcomers}. Mentoring is a role often connected to this stereotype. We see an increase in women mentors (4.16\%), but there is a bigger increase among men mentors (7.54\%). This possibly reflects an increased appreciation of the importance of mentoring activities among contributors from Apache projects, a trend that can help attract and retain newcomers. 

Despite the progress on these two fronts, more must be done. We saw a decline in gender minority respondents' engagement with project development decision-making (2.41\%). This reflects the need to promote women to leadership positions, which is considered an effective solution to foster diversity \cite{bosu2019diversity}. We recommend that the PMC and the board members actively engage those in gender minorities in their decision-making at the project and foundation levels. This strategy was also cited by Trinkenreich et al. \cite{trinkenreich2022women}.

\begin{mdframed}[roundcorner=10pt,nobreak=true]
\textbf{Observation 1:} More respondents from gender minorities are engaging in both code and non-code contributions, but there is an increasing gap in participation in decision-making activities.
\end{mdframed}

\textbf{Seniority:} There is an increase in engagement across all listed activities among non-newcomer respondents (seniority$>$1 year), ranging from 1\% to 6\% increase. The proportion of newcomer respondents who engage in code-related activities increased by 19.87\%, which may allude to a healthy trend of Apache projects being able to attract newcomers and ensure they succeed in their contributions. Additionally, newcomer respondents are becoming increasingly involved in decision-making (7.68\%) and mentoring others (13.04\%). OSS contributors frequently contribute to many projects and shift between them ~\cite{jergensen2011onion}. In such instances, they may acquire the necessary skills and experience from other sources and continue to share their knowledge while contributing to OSS, regardless of their seniority within the project. Such mentoring activities have been defined as implicit mentoring, where contributors guide each other in everyday OSS activities such as code review \cite{feng2022case}. This implies that OSS communities are not only recruiting newcomers but also newcomers feel engaged and become community members by implicitly mentoring other contributors.

One point of concern could be the decrease in the percentage of newcomers participating in creating or maintaining documentation (4.64\%), along with a 4.56\% increase among senior contributors. On the one hand, these results show that senior contributors are engaging more in non-code contributions, which is good for the sustainability of the community. On the other hand, it can become a cause of concern if the documentation ends up being written from the point of view of senior contributors, making it harder for newcomers. The literature shows that a lack of documentation and roadmap to participation and ambiguous and outdated documentation impedes newcomers \cite{steinmacher2014hard, steinmacher2015social}. We recommend that OSS projects review their documentation to ensure newcomers' documentation needs are met. Involving newcomers in these activities can help both to engage the newcomers and to ensure the documentation is accessible for those outside the project.


\begin{mdframed}[roundcorner=10pt,nobreak=true]
\textbf{Observation 2:} Newcomers engage more actively in code contribution, code review, mentoring other contributors, and decision-making processes. However, they tend to be less involved in community organizations and documentation-related activities.
\end{mdframed}

\textbf{English proficiency:} English is the most common language in  Apache projects. Therefore, we use it as a diversity attribute, as contributors who are not proficient in English can face additional barriers to their efforts \cite{guizani2021long, balali2018newcomers}. 

Contributors who considered themselves not proficient in English showed higher engagement across all activities (except, as already discussed, in the community organization). The largest increase (25.56\%) was in mentoring others. This can be a side effect of contributors feeling more comfortable seeking help from those with similar English proficiency levels. However, as this is an anonymous survey, we cannot associate the demographics of mentors with that of mentees, which can be a future investigation. We found that contributors who are proficient in English did not reveal significant variations in activity engagement between the two surveys (less than 6 percent).

Respondents not comfortable using English were also more involved in creating/maintaining documentation (11.13\%) and decision-making (1.67\%). This implies that the OSS community is becoming more inclusive of contributors whose native language is not English.

\begin{mdframed}[roundcorner=10pt,nobreak=true]
\textbf{Observation 3:} Our results show that English proficiency seems to be becoming less of a barrier to participation in project activities, as non-native English-speaking respondents are increasingly involved in creating and maintaining project documentation and mentoring other contributors.
\end{mdframed}

\textbf{Geo-location:} A majority of Apache projects that successfully incubate are from North America and Europe, as per the Apache D\&I committee. Apache projects are actively looking at strategies to promote the incubation success of projects from Asian countries, especially China. In addition to the incubation process challenges that contributors face, such as ``compliance and project ascension to top level'' \cite{guizani2021long}, language barriers, cultural differences, and communication styles can impede contributors from non-western-centric countries \cite{chung2013time, steinmacher2016overcoming, guizani2021long}.

We see an increase in the proportion of non-Western respondents engaged in the OSS activities, including being community organizers (an increase of 17.98\%, the largest increase in this activity across all demographic attributes). Most other activities also had double-digit improvements; a 23.9\% increase in being mentors, a 19.85\% increase in participating in decision-making, and an 18.44\% improvement in creating/maintaining documents. These numbers show that Apache projects are becoming more inclusive of contributors from non-western countries. The smallest improvement was in contributing to code/reviewing (7.37\%). These results show greater improvements in non-code related activities, which can provide a good pathway toward more code-related activities~\cite{trinkenreich2020hidden}. 

\begin{mdframed}[roundcorner=10pt,nobreak=true]
\textbf{Observation 4}: There is increased participation from respondents from non-western countries across all activities.
\end{mdframed}

\begin{table*}[tbp]
\caption{Comparison of respondents facing challenges frequently between 2020 and 2022.}
\centering
\resizebox{7in}{!}{
\begin{tabular}{lll|ll|ll|ll|ll|ll|ll|ll}
\multicolumn{1}{c}{\textbf{Questions}} & \multicolumn{2}{c|}{\textbf{Man}} & \multicolumn{2}{c|}{\textbf{Minority gender}} & \multicolumn{2}{c|}{\textbf{Newcomer}} & \multicolumn{2}{c|}{\textbf{Not newcomer}} & \multicolumn{2}{c|}{\textbf{\begin{tabular}[c]{@{}c@{}}Comfortable using\\ English\end{tabular}}} & \multicolumn{2}{c|}{\textbf{\begin{tabular}[c]{@{}c@{}}Not comfortable using \\ English\end{tabular}}} & \multicolumn{2}{c|}{\textbf{\begin{tabular}[c]{@{}c@{}}From western\\ countries\end{tabular}}} & \multicolumn{2}{c}{\textbf{\begin{tabular}[c]{@{}c@{}}Not from western \\ countries\end{tabular}}} \\
\rowcolor[HTML]{EFEFEF} 
Challenges & \reddown & {\color[HTML]{C70039} -3.04\%} & \reddown & {\color[HTML]{C70039} -6.97\%} & \reddown & {\color[HTML]{C70039} -8.93\%} & \reddown & {\color[HTML]{C70039} -2.99\%} & \reddown & {\color[HTML]{C70039} -2.87\%} & \reddown & {\color[HTML]{C70039} -4.61\%} & \reddown & {\color[HTML]{C70039} -3.17\%} & \reddown & {\color[HTML]{C70039} -4.60\%}
\end{tabular}}
\label{tab:challenge}
\end{table*}

\subsection{Challenges to contributing}

Past works have shown that respondents from non-majority groups face challenges in contributing to OSS \cite{steinmacher2016overcoming, terrell2017gender, singh2019open, prana2021including}. This is true for newcomers and existing contributors. Guizani et al. \cite{guizani2021long} categorized the challenges that contributors face into three groups: (1) technical challenges, which are related to technical hurdles with the project code, its infrastructure, or lack of documentation; (2) social challenges, which are related to the communication styles and (unwelcoming) project culture; and (3) process-related challenges, which are related to navigating the contribution process, getting started, and licenses. Here we build on this categorization and analyze if there was any improvement (or worsening situation) to the frequency of challenges contributors face, disaggregated by the demographic aspects. 

Overall, there was a reduction in the percentage of respondents who mentioned frequently facing challenges. \Cref{tab:challenge} shows that fewer respondents across all different demographic groups reported frequently facing challenges. Among the majority groups, the percentage reduction ranged between about 3\% to 5\%; 2.99\% fewer seniors and 4.60\% respondents from western countries reported facing challenges frequently.

The reduction in proportions was higher for the minority groups. The decline was highest among respondents from newcomers (8.93\%) followed by gender minority groups (6.97\%). This signals that the programs and initiatives to improve gender diversity and attract newcomers to Apache projects have a positive impact. The proportion of respondents from non-western countries and lower English proficiency reduced too, but in the 4\% range. We recommend OSS projects review its documentation and communication processes to further reduce challenges faced by contributors who are non-native English speakers. 

\begin{mdframed}[roundcorner=10pt,nobreak=true] 
\textbf{Observation 5}: The challenges faced by respondents have reduced across all demographic groups; the reduction is higher for those in minority groups.
\end{mdframed}

\begin{figure}[!tbp]
\centering
\includegraphics[width=3.5 in]{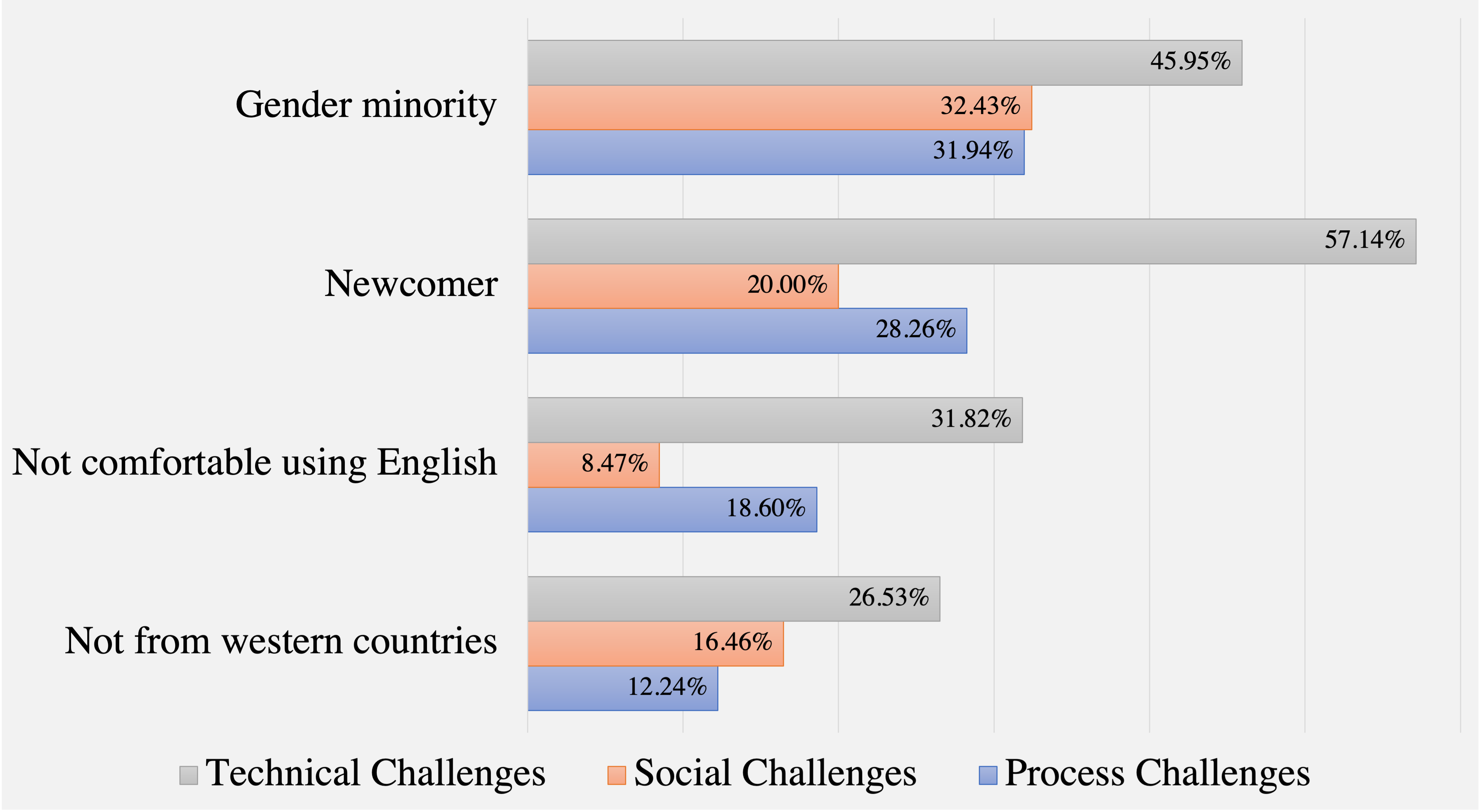}
\caption{The three types of challenges that contributors face disaggregated by demographics in the 2022 survey.}
\label{fig:challenges}
\end{figure}

Next, we delve deeper into the types of challenges that contributors from minority groups are still facing. The goal of this pulse check on the types of challenges is to help the communities identify challenge ``hot spots", so they can prioritize their inclusion strategies.

\Cref{fig:challenges} shows three types of challenges that contributors from minority demographic groups faced, according to the 2022 survey. While the 2020 survey asked for challenges in an open-ended question, the 2022 survey used Likert scale questions on specific types of challenges based on Guizani et al.'s conceptual model of the challenges faced by contributors \cite{guizani2021long} (see Q6 in \cref{tab:survey}).

\textit{Technical challenges} were the most prevalent across all categories. The highest incidence is for newcomers (57.14\% respondents indicated that they ``often'' confront technical challenges). While newcomers facing technical challenges is to be expected, since they have to learn the project code and its infrastructure, a 57\% proportion is concerning. This was followed by gender-minority groups (45.95\%), which can be a result of OSS projects and documents not supporting the cognitive styles favored by women \cite{mendez2018open}. This was followed by respondents not proficient in English (31.82\%) and those from not western countries (26.53\%). These results indicate that perhaps the documentation can be more language-inclusive (e.g., not using idioms, jargon) to make it easier to comprehend. We recommend that OSS projects perform an inclusivity evaluation of its project and documentation, perhaps using the GenderMag method as done by Guizani et al. \cite{guizani2022debug}.

\textit{Social challenges} were more prominently reported by gender minorities (32.43\%) and newcomers (20\%). Social challenges to participation arising from non-inclusive communication, toxic culture, and stereotyping \cite{trinkenreich2022women, balali2018newcomers} are often barriers to participation by those in underrepresented groups. Similarly, when it comes to \textit{process issues}, we found that they were more frequent for gender minority  (31.94\%) newcomers (28.26\%) respondents. Potential explanations may be related to a lack of project orientation, a lack of up-to-date documentation, and reception problems resulting from bias \cite{steinmacher2014hard, trinkenreich2022women}.

Our findings indicate that while fewer respondents reported facing challenges frequently, OSS projects still need to strive to remove gender biases and stereotypes to create an inclusive and welcoming environment.

\subsection{Perceptions about the state of D\&I}

Feeling represented and valued by the community is important to being productive and satisfied~\cite{lim2008job,trinkenreich2023sense}. Past work found that certain factors can impede the feeling of being represented, such as the lack of interpersonal relationships in the community~\cite{tinto1987leaving} and the perception that one's voice is lost in an environment where the loudest voice prevails~\cite{nafus2012patches}. Therefore, here we analyze contributors' perceptions of their ability to contribute to Apache projects. 

As before, we compare the survey results regarding the perception of D\&I between 2020 and 2022 surveys as shown in \Cref{tab:statement}. 

\begin{table*}[tbp]
\caption{Comparison of respondents' agreement on the perceptions of D\&I between the years 2020 and 2022.}
\centering
\resizebox{\textwidth}{!}{
\begin{tabular}{clll|llll|ll|ll|ll|ll|ll}
\textbf{Perception of} & \multicolumn{1}{c}{\textbf{Questions}} & \multicolumn{2}{c|}{\textbf{Man}} & \multicolumn{2}{c|}{\textbf{Minority gender}} & \multicolumn{2}{c|}{\textbf{Newcomer}} & \multicolumn{2}{c|}{\textbf{Not newcomer}} & \multicolumn{2}{c|}{\textbf{\begin{tabular}[c]{@{}c@{}}Comfortable using\\ English\end{tabular}}} & \multicolumn{2}{c|}{\textbf{\begin{tabular}[c]{@{}c@{}}Not comfortable using \\ English\end{tabular}}} & \multicolumn{2}{c|}{\textbf{\begin{tabular}[c]{@{}c@{}}From western\\ countries\end{tabular}}} & \multicolumn{2}{c}{\textbf{\begin{tabular}[c]{@{}c@{}}Not from western \\ countries\end{tabular}}} \\ \hline
 & \cellcolor[HTML]{EFEFEF}\begin{tabular}[c]{@{}l@{}}Other members of the project \\ see me as a parental figure\end{tabular} & \cellcolor[HTML]{EFEFEF}\greenup & \cellcolor[HTML]{EFEFEF}{\color[HTML]{228B22} 10.15\%} & \cellcolor[HTML]{EFEFEF}\greenup & \multicolumn{1}{l|}{\cellcolor[HTML]{EFEFEF}{\color[HTML]{228B22} 14.13\%}} & \cellcolor[HTML]{EFEFEF}\greenup & \cellcolor[HTML]{EFEFEF}{\color[HTML]{228B22} 10.92\%} & \cellcolor[HTML]{EFEFEF}\greenup & \cellcolor[HTML]{EFEFEF}{\color[HTML]{228B22} 9.08\%} & \cellcolor[HTML]{EFEFEF}\greenup & \cellcolor[HTML]{EFEFEF}{\color[HTML]{228B22} 10.44\%} & \cellcolor[HTML]{EFEFEF}\greenup & \cellcolor[HTML]{EFEFEF}{\color[HTML]{228B22} 8.70\%} & \cellcolor[HTML]{EFEFEF}\greenup & \cellcolor[HTML]{EFEFEF}{\color[HTML]{228B22} 10.57\%} & \cellcolor[HTML]{EFEFEF}\greenup & \cellcolor[HTML]{EFEFEF}{\color[HTML]{228B22} 10.91\%} \\
 & \begin{tabular}[c]{@{}l@{}}I am expected to take care \\ of other  members of the\\ project more so than is usual\end{tabular} & \greenup & {\color[HTML]{228B22} 4.27\%} & \reddown & \multicolumn{1}{l|}{{\color[HTML]{C70039} -15.63\%}} & \greenup & {\color[HTML]{228B22} 8.16\%} & \greenup & {\color[HTML]{228B22} 1.23\%} & \greenup & {\color[HTML]{228B22} 2.34\%} & \greenup & {\color[HTML]{228B22} 8.41\%} & \greenup & {\color[HTML]{228B22} 2.37\%} & \reddown & {\color[HTML]{C70039} -1.06\%} \\
\multirow{-3}{*}{\textbf{\begin{tabular}[c]{@{}c@{}}Role \\ stereotyping\end{tabular}}} & \cellcolor[HTML]{EFEFEF}\begin{tabular}[c]{@{}l@{}}I feel some members of the\\ community are  patronizing \\ to me\end{tabular} & \cellcolor[HTML]{EFEFEF}\reddown & \cellcolor[HTML]{EFEFEF}{\color[HTML]{C70039} -2.33\%} & \cellcolor[HTML]{EFEFEF}\reddown & \multicolumn{1}{l|}{\cellcolor[HTML]{EFEFEF}{\color[HTML]{C70039} -12.28\%}} & \cellcolor[HTML]{EFEFEF}\reddown & \cellcolor[HTML]{EFEFEF}{\color[HTML]{C70039} -4.42\%} & \cellcolor[HTML]{EFEFEF}\reddown & \cellcolor[HTML]{EFEFEF}{\color[HTML]{C70039} -2.82\%} & \cellcolor[HTML]{EFEFEF}\reddown & \cellcolor[HTML]{EFEFEF}{\color[HTML]{C70039} -1.85\%} & \cellcolor[HTML]{EFEFEF}\reddown & \cellcolor[HTML]{EFEFEF}{\color[HTML]{C70039} -13.55\%} & \cellcolor[HTML]{EFEFEF}\reddown & \cellcolor[HTML]{EFEFEF}{\color[HTML]{C70039} -2.58\%} & \cellcolor[HTML]{EFEFEF}\reddown & \cellcolor[HTML]{EFEFEF}{\color[HTML]{C70039} -5.15\%} \\ \hline
 & \begin{tabular}[c]{@{}l@{}}I have an equal chance to\\ get contributions accepted\end{tabular} & \greenup & {\color[HTML]{228B22} 3.76\%} & \greenup & \multicolumn{1}{l|}{{\color[HTML]{228B22} 8.97\%}} & \greenup & {\color[HTML]{228B22} 9.95\%} & \greenup & {\color[HTML]{228B22} 1.45\%} & \greenup & {\color[HTML]{228B22} 0.20\%} & \greenup & {\color[HTML]{228B22} 19.12\%} & \greenup & {\color[HTML]{228B22} 0.93\%} & \greenup & {\color[HTML]{228B22} 4.78\%} \\

 & \cellcolor[HTML]{EFEFEF}\begin{tabular}[c]{@{}l@{}}Nothing keeps me from\\ contributing to the project\end{tabular} & \cellcolor[HTML]{EFEFEF}\reddown & \cellcolor[HTML]{EFEFEF}{\color[HTML]{C70039} -0.06\%} & \cellcolor[HTML]{EFEFEF}\greenup & \multicolumn{1}{l|}{\cellcolor[HTML]{EFEFEF}{\color[HTML]{228B22} 39.43\%}} & \cellcolor[HTML]{EFEFEF}\greenup & \cellcolor[HTML]{EFEFEF}{\color[HTML]{228B22} 10.45\%} & \cellcolor[HTML]{EFEFEF}\greenup & \cellcolor[HTML]{EFEFEF}{\color[HTML]{228B22} 0.70\%} & \cellcolor[HTML]{EFEFEF}\greenup & \cellcolor[HTML]{EFEFEF}{\color[HTML]{228B22} 0.57\%} & \cellcolor[HTML]{EFEFEF}\greenup & \cellcolor[HTML]{EFEFEF}{\color[HTML]{228B22} 14.11\%} & \cellcolor[HTML]{EFEFEF}\greenup & \cellcolor[HTML]{EFEFEF}{\color[HTML]{228B22} 0.77\%} & \cellcolor[HTML]{EFEFEF}\greenup & \cellcolor[HTML]{EFEFEF}{\color[HTML]{228B22} 2.41\%} \\

 & \begin{tabular}[c]{@{}l@{}}I have a solid network of\\ open source peers\end{tabular} & \reddown & {\color[HTML]{C70039} -2.09\%} & \greenup & \multicolumn{1}{l|}{{\color[HTML]{228B22} 18.08\%}} & \greenup & {\color[HTML]{228B22} 1.97\%} & \reddown & {\color[HTML]{C70039} -2.78\%} & \reddown & {\color[HTML]{C70039} -2.50\%} & \greenup & {\color[HTML]{228B22} 11.90\%} & \reddown & {\color[HTML]{C70039} -2.83\%} & \greenup & {\color[HTML]{228B22} 13.27\%} \\

 & \cellcolor[HTML]{EFEFEF}\begin{tabular}[c]{@{}l@{}}It was easy to find a mentor\\ with whom I  felt comfortable\end{tabular} & \cellcolor[HTML]{EFEFEF}\reddown & \cellcolor[HTML]{EFEFEF}{\color[HTML]{C70039} -5.86\%} & \cellcolor[HTML]{EFEFEF}\greenup & \multicolumn{1}{l|}{\cellcolor[HTML]{EFEFEF}{\color[HTML]{228B22} 11.76\%}} & \cellcolor[HTML]{EFEFEF}\reddown & \cellcolor[HTML]{EFEFEF}{\color[HTML]{C70039} -7.18\%} & \cellcolor[HTML]{EFEFEF}\reddown & \cellcolor[HTML]{EFEFEF}{\color[HTML]{C70039} -4.29\%} & \cellcolor[HTML]{EFEFEF}\reddown & \cellcolor[HTML]{EFEFEF}{\color[HTML]{C70039} -4.38\%} & \cellcolor[HTML]{EFEFEF}\reddown & \cellcolor[HTML]{EFEFEF}{\color[HTML]{C70039} -7.83\%} & \cellcolor[HTML]{EFEFEF}\reddown & \cellcolor[HTML]{EFEFEF}{\color[HTML]{C70039} -5.63\%} & \cellcolor[HTML]{EFEFEF}\reddown & \cellcolor[HTML]{EFEFEF}{\color[HTML]{C70039} -1.88\%} \\
\multirow{-5}{*}{\textbf{\begin{tabular}[c]{@{}c@{}}Ability to \\ contribute\end{tabular}}} & \begin{tabular}[c]{@{}l@{}}I have a hard time following\\ discussions because of technical \\ jargon\end{tabular} &  & {*} &  & \multicolumn{1}{l|}{*} & \greenup & {\color[HTML]{228B22} 3.04\%} & \reddown & {\color[HTML]{C70039} -1.65\%} & \reddown & {\color[HTML]{C70039} -1.17\%} & \reddown & {\color[HTML]{C70039} -0.26\%} & \reddown & {\color[HTML]{C70039} -1.53\%} & \greenup & {\color[HTML]{228B22} 0.08\%} \\ \hline
 & \cellcolor[HTML]{EFEFEF}\begin{tabular}[c]{@{}l@{}}The PMC represents a diverse \\ set of people\end{tabular} & \cellcolor[HTML]{EFEFEF}\greenup & \cellcolor[HTML]{EFEFEF}{\color[HTML]{228B22} 1.56\%} & \cellcolor[HTML]{EFEFEF}\reddown & \multicolumn{1}{l|}{\cellcolor[HTML]{EFEFEF}{\color[HTML]{C70039} -22.22\%}} & \cellcolor[HTML]{EFEFEF}\reddown & \cellcolor[HTML]{EFEFEF}{\color[HTML]{C70039} -7.69\%} & \cellcolor[HTML]{EFEFEF}\greenup & \cellcolor[HTML]{EFEFEF}{\color[HTML]{228B22} 1.15\%} & \cellcolor[HTML]{EFEFEF}\greenup & \cellcolor[HTML]{EFEFEF}{\color[HTML]{228B22} 0.21\%} & \cellcolor[HTML]{EFEFEF}\greenup & \cellcolor[HTML]{EFEFEF}{\color[HTML]{228B22} 13.54\%} & \cellcolor[HTML]{EFEFEF}\reddown & \cellcolor[HTML]{EFEFEF}{\color[HTML]{C70039} -2.58\%} & \cellcolor[HTML]{EFEFEF}\greenup & \cellcolor[HTML]{EFEFEF}{\color[HTML]{228B22} 14.44\%} \\
\multirow{-2}{*}{\textbf{\begin{tabular}[c]{@{}c@{}}Being \\ represented\end{tabular}}} & \begin{tabular}[c]{@{}l@{}}I feel represented in the decision \\ making  group\end{tabular} & \greenup & {\color[HTML]{228B22} 4.83\%} & \reddown & \multicolumn{1}{l|}{{\color[HTML]{C70039} -16.40\%}} & \greenup & {\color[HTML]{228B22} 7.28\%} & \greenup & {\color[HTML]{228B22} 2.36\%} & \greenup & {\color[HTML]{228B22} 1.22\%} & \greenup & {\color[HTML]{228B22} 8.53\%} & \greenup & {\color[HTML]{228B22} 1.40\%} & \greenup & {\color[HTML]{228B22} 13.75\%} \\ \hline
 & \cellcolor[HTML]{EFEFEF}\begin{tabular}[c]{@{}l@{}}I was made aware of the code\\ of conduct and how to report \\ violations\end{tabular} & \cellcolor[HTML]{EFEFEF}\greenup & \cellcolor[HTML]{EFEFEF}{\color[HTML]{228B22} 7.83\%} & \cellcolor[HTML]{EFEFEF}\greenup & \multicolumn{1}{l|}{\cellcolor[HTML]{EFEFEF}{\color[HTML]{228B22} 7.89\%}} & \cellcolor[HTML]{EFEFEF}\reddown & \cellcolor[HTML]{EFEFEF}{\color[HTML]{C70039} -13.15\%} & \cellcolor[HTML]{EFEFEF}\greenup & \cellcolor[HTML]{EFEFEF}{\color[HTML]{228B22} 9.47\%} & \cellcolor[HTML]{EFEFEF}\greenup & \cellcolor[HTML]{EFEFEF}{\color[HTML]{228B22} 7.98\%} & \cellcolor[HTML]{EFEFEF}\greenup & \cellcolor[HTML]{EFEFEF}{\color[HTML]{228B22} 8.88\%} & \cellcolor[HTML]{EFEFEF}\greenup & \cellcolor[HTML]{EFEFEF}{\color[HTML]{228B22} 5.47\%} & \cellcolor[HTML]{EFEFEF}\greenup & \cellcolor[HTML]{EFEFEF}{\color[HTML]{228B22} 14.90\%} \\
\multirow{-2}{*}{\textbf{\begin{tabular}[c]{@{}c@{}}The code\\  of conduct\end{tabular}}} & \begin{tabular}[c]{@{}l@{}}I felt safer and more empowered \\ to fully  participate in this project \\ because it followed the code of \\ conduct\end{tabular} & \greenup & {\color[HTML]{228B22} 6.69\%} & \greenup & \multicolumn{1}{l|}{{\color[HTML]{228B22} 11.90\%}} & \greenup & {\color[HTML]{228B22} 3.40\%} & \greenup & {\color[HTML]{228B22} 8.14\%} & \greenup & {\color[HTML]{228B22} 8.83\%} & \reddown & {\color[HTML]{C70039} -1.01\%} & \greenup & {\color[HTML]{228B22} 5.44\%} & \greenup & {\color[HTML]{228B22} 6.11\%} \\ \hline
\multicolumn{1}{l}{} & \multicolumn{17}{l}{\cellcolor[HTML]{EFEFEF}* We posit gender should not impact following technical discussion, thus, we dont present it in table. For reference, the breakdown is: man: -1.54\%;  minority-gender: -2.19\%.}
\end{tabular}}
\label{tab:statement}
\end{table*}

\textbf{Role stereotyping:} The proportion of respondents who perceived that they were seen as ``parental figures" and were ``expected to take care of others \textit{more than usual}'' increased for all demographics, except for those from non-gender minority for the ``taking care" question. An explanation of this trend could be that with the reduction in community organization or events along with an increase in the proportion of respondents contributing to code and non-code artifacts, there is an increased need for mentoring. Every demographic (majority and minority) reported a higher proportion of mentoring others. This might result in respondents feeling that they are perceived as parental figures and have to take care of others. Interestingly, gender minority respondents had a reduction (15.63\%) in feeling that they were expected to take care of others more than usual. It might be that because other demographics have stepped up that non-minority contributors feel a reduction, or it might be that these contributors have a different expectation of what is considered ``usual". We hope it is the former.

There is a positive trend in the community being more welcoming. The proportion of respondents who felt members were ``patronizing" reduced across all demographics, for both majority and minority. The reduction was in double digits for those in minority-gender (12.28\%) and not English proficient (13.55\%) groups. We hope this trend reflects OSS is becoming more inclusive and respectful of its members, including those traditionally underrepresented.

\begin{mdframed}[roundcorner=10pt,nobreak=true]
\textbf{Observation 6}: Respondents across all demographics feel they are sought out as parental figures, while fewer proportions feel patronized.
\end{mdframed}

\textbf{Ability to contribute:} Biases against gender minorities, language proficiency, and cultures can be seen as hurdles to contributing to OSS. Results indicate that contribution barriers have been reduced for all the analyzed demographic factors, with a higher proportion of contributors feeling they have an ``equal chance" of getting contributions accepted, especially minority-gender (8.97\%) and not English-proficient (19.12\%) contributors. We also observed a higher proportion of newcomers who feel positive (9.95\%) about having an ``equal chance" of getting contributions accepted. Similar improvements in sentiments are seen for the ``nothing keeps me from contributing" question (39.43\% for gender minorities and 14.11\% for non-English proficients). This positive trend can help Apache projects attract and retain contributors.

When considering ``network of peers" and ``finding a mentor they are comfortable with", we see mixed results. A higher proportion of contributors from gender minorities feel they have a good peer network and mentors they are comfortable with. All other minority demographics reported an improvement in having a solid network of peers but a reduction in the ability to find mentors they are comfortable with.  

These results reflect a positive trend of contributors having a network of peers, on who they can depend. This could also be a reason for us seeing a higher proportion of contributors performing mentoring activities (\Cref{tab:engagement}) and those who feel they are seen as parental figures. However, it appears that finding a desired mentor remains challenging. Another reason for this dichotomy (more contributors are mentoring, but fewer respondents are finding mentors they feel comfortable with) could be a result of the mentorship programs. Successful mentor-mentee relationships often occur organically based on topics of interest or alignment of personality and career goals. Formal mentorship programs can often feel forced or misaligned based on differences between mentor-mentee working styles and goals~\cite{balali2018newcomers}. We recommend that OSS community keep developing mentorship programs and investigate strategies to facilitate and encourage informal mentoring.

When considering following technical discussions, there is some improvement for newcomers (3.04\%), which indicates either newcomers are getting more familiar with technical jargon or that the discussions are using less jargon and being more inclusive. The differences for the rest of the demographics are marginal. 

\begin{mdframed}[roundcorner=10pt,nobreak=true]
\textbf{Observation 7}: Respondents agreeing on the equal chance to contribute has increased across all demographic factors.  However, when it comes to finding an ideal mentor, there is less agreement among all groups except for those who identify as a minority gender. 
\end{mdframed}

\textbf{Perception of being represented:}
There has been a significant impetus in increasing awareness of gender diversity, both in research and in practice. In fact, 12 out of 355 OSS websites have a ``women-only'' section \cite{Lee.Carver:2019}. Our results show that there is still work to be done at the leadership level. A higher proportion of those in gender minorities (22.22\%) and newcomers (7.69\%) disagreed about PMC representing a diverse set of people. Fewer proportion of those in gender minorities felt that they were represented in the decision-making group. These results imply that OSS needs to actively mentor and promote those in gender minorities to leadership positions to improve their representation.

It is heartening to see that those not in western countries and not proficient in English feel more representation---both in the PMC makeup and in decision-making groups.

\begin{mdframed}[roundcorner=10pt,nobreak=true]
\textbf{Observation 8}: Gender minority and newcomer respondents were less likely to concur that they felt represented by PMC and decision-making.
\end{mdframed}

\textbf{Code of conduct:}
A code of conduct that outlines the expected behaviors of its members can assist in creating a more pleasant and supportive social environment \cite{CoC}. Creating a code of conduct is one of the most widely used D\&I efforts in OSS \cite{guizani2022perceptions,trinkenreich2022women}. Apache projects rely primarily on English-language content, and respondents who are not proficient in English and are not from western countries may encounter language barriers. Due to differences in culture and geography, rules and procedures such as the code of conduct may lose attention and create additional communication hurdles \cite{li2021code}.

Our results indicate that the awareness of the code of conduct has increased by 8.88\% for respondents who are not comfortable using English and 14.90\% for respondents who are not from western countries. Similarly, respondents from non-Western countries and non-native English speakers feel more empowered to participate because their projects follow the code of conduct. 

Nonetheless, we found that the proportion of newcomers and those in gender minorities who were aware of the code of conduct decreased (13.15\% for newcomers). These trends might mean that more contributors are unaware of how to react to unwelcoming interactions, and this might be why we saw an increase in perceptions of ``members are patronizing to me'' for these subgroups. This lack of awareness can be a result of the reduced number of events where respondents from these groups could interact informally about an unwelcoming environment and talk about action plans.

On the flip side, more respondents from these groups (arguably, when they were aware of the code of conduct) felt empowered to participate fully in the project when it followed its code of conduct. This could be because increasingly code-of-conduct documents are being written not simply as an ``aspirational" document, but with actionable steps that need to be taken when someone violates the established practices and rules.

Our recommendation is twofold. First, the code-of-conduct document should be prominently listed in projects, perhaps even as a separate tab. Additionally, onboarding documents, training, and mentorship programs should emphasize the code of conduct. Second, projects should follow their code of conduct, ensuring that those who do not follow it are appropriately penalized.

\begin{mdframed}[roundcorner=10pt,nobreak=true]
\textbf{Observation 9}: When projects follow their code-of-conduct respondents feel empowered to participate.
\end{mdframed}

\section{Discussion}

Longitudinal studies are needed to understand the evolution of the state of D\&I and to discern if diversity initiatives have an impact. As part of our longitudinal study sponsored by the Apache diversity project, here we provide a pulse check of the state of D\&I across two surveys (across a three-year time period). Our results show improvements in the state of D\&I, which indicates the Apache projects are on their way to being more inclusive. In the following, we discuss where the results indicate improvements and where they indicate that progress is still needed.

\subsection{The state of D\&I is Improving}

Our findings indicate that the hurdles arising because of biases and stereotypes are diminishing, particularly for underrepresented groups such as respondents of gender-minority and respondents with limited English proficiency. We posit that this is due to increased awareness of the importance of D\&I among the public as well as the Apache's efforts to make the project more inclusive, such as its mentoring program and its efforts to create a welcoming workplace \cite{EDIgroup}. 

\textbf{Gender inclusive:} A key impact of the diversity gap in OSS and role stereotype is disparity in becoming a contributor and having contributions accepted~\cite{guizani2022debug}. Our results show that from 2020 to 2022 there is a marked improvement (30.52\%) among gender-minority respondents in making code-related contributions as well as improvements in their perception about their ability to contribute. 6.97\% fewer respondents in this class reported facing challenges frequently. Fewer participants also reported facing role stereotyping. 

One form of role stereotyping is that women are considered to be warm and nurturing and seen as parental figures. They end up taking a larger share of community organizing and mentoring ~\cite{trinkenreich2020hidden}. However, our survey indicates that more men are taking up mentoring activities and seen as parental figures.  


\textbf{Geo-location, English Proficiency:} Results indicate that respondents who are not from western countries and are not proficient in English are increasingly engaging in the majority of the activities. Further, the challenges and barriers that arise because of these demographic aspects are on the decline. This increase in their activities aligns with their perception of being able to contribute to OSS. As with gender-minority, fewer respondents in these categories reported facing challenges in 2022 as compared to 2020. More respondents in these categories also perceived that the PMC is diverse and they are represented in decision making.

\subsection{Progress is still needed}

\textbf{Developing a gender inclusive community:} There still exists a significant gap between the number of respondents who identify as men versus gender-minority respondents. Additionally, many biases and gender-barriers are entrenched in OSS, as men have (and still) dominate OSS community ~\cite{trinkenreich2022women}. Improving the state of D\&I is therefore a long-term goal. While there are improvements in OSS projects being gender inclusive, there is still more to be done. OSS community should continue in their outreach and awareness program to attract more contributors who are from gender minority.
In particular, OSS could be intentional in promoting and mentoring women to leadership roles, which will also help attract and retain newcomers from underrepresented groups~\cite{bosu2019diversity}.

\textbf{Welcoming newcomers to OSS:} How to appropriately and effectively onboard newcomers is always an ongoing challenge in OSS. OSS communities are trying hard to utilize mentoring programs for training future newcomers, such as Google Summer of Code \cite{googlecode,silva2020theory} and the Linux mentoring program \cite{linux}. Indeed, regardless of their demographic backgrounds, we found that respondents are mentoring other contributions more frequently. However, fewer respondents concurred that locating a desirable mentor is easy. 
Potential causes for this problem could be time zone conflicts, mentor-mentee interest mismatches, or negative sentiments about receiving negative feedback \cite{balali2018newcomers}.   

We believe there are two ways to mitigate this problem. First, promote and acknowledge mentoring activities. For example, Feng et al. \cite{feng2022case} presented a mechanism to identify implicit mentoring, where topical mentoring is done through code reviews. In their study,  over 90\% of their survey respondents found such mentoring to be more effective and beneficial for both mentors and mentees as compared to formal mentoring programs. OSS projects should, therefore, acknowledge and highlight such implicit mentors. Second, in addition to formal mentoring programs, OSS community should investigate organizing informal mentoring events where mentors and mentees can organically meet and create mentorship relations. 


\textbf{Intersectionality and Multidimensionality:}
Our analysis revealed that the state of diversity in OSS is improving across multiple attributes. However, to further enhance the state of D\&I in OSS, it is essential to consider intersectionality, where contributors can be identified by multiple attributes, such as a woman contributor who is also a newcomer and lacks English confidence. Future research should investigate how these factors interact and impact contributors' daily OSS activities. For example, In what ways do gender minority senior contributors impact mentoring activities? How do gender minority contributors who are not confident in English experience challenges, and which D\&I interaction patterns are more likely to face challenges? Additionally, do perceptions associated with both seniority and English proficiency exacerbate their feeling of being patronized?

As diversity is a multidimensional concept, a unidimensional perspective is insufficient to comprehend the state of D\&I in OSS community in its entirety. To investigate the interplay between various D\&I attributes, a multidimensional perspective is necessary. Thus, the community can more effectively adapt and develop strategies to promote diversity and inclusion. Moreover, our survey design  and research approach can serve as a useful starting point for future research to explore how to tailor strategies to promote inclusivity in OSS for different groups of contributors with diverse intersectional identities.
\section{Threats to Validity}

As with all empirical research, there are threats to validity associated with our study, despite our best efforts to mitigate them.

\textbf{External:} We surveyed only ASF contributors and our research findings may not directly generalize to other OSS communities. On the other hand, ASF is very diverse in terms of contributions. It hosts 300+ projects in a wide variety of fields and has thousands of contributors from all over the world.

\textbf{Internal:} 
Internal validity threats might arise from a biased sampling of the ASF contributors. Although we deployed the surveys widely and received 600+ and 400+ responses, self-selection bias could have affected our results. 

\textbf{Construct:} Inaccurate measurements could result from us asking the wrong questions in our survey or interviews. To mitigate this threat, we utilized survey questions from existing studies and ASF internal survey recommended by the ASF D\&I committee \cite{2016ASFSurvey, Lee.Carver:2019}. Moreover, the final questionnaires we used for our longitudinal study were developed in collaboration with sixteen experienced ASF contributors from the D\&I committee.

\textbf{Conclusion:} This is a preliminary approach to analyzing the D\&I in OSS over time. Our longitudinal study consists of only two surveys conducted over two years. Survival bias may result in an ``optimistic'' perspective. As a result of the difference in the number of respondents from each background attribute, we may have obtained fewer results when observing proportional differences between the two survey outcomes. In addition, due to the fact that this is a preliminary study with only two years of survey data and we have an under-representation of minorities, statistical analysis can be biased. We only present the proportional differences we noticed; in the future, when we acquire more survey data, a more in-depth analysis will be required.

\section{Conclusion and Future work}

In this study, we conducted a longitudinal approach to investigate the evolution of D\&I within a large OSS organization by conducting two cross-sectional surveys in Apache projects. Our findings show that the obstacles and challenges resulting from diverse backgrounds and experiences are on the path to improvement, particularly for contributors from groups that are traditionally considered underrepresented.

To continue promoting an inclusive community and organizing D\&I activities, it is necessary to keep informing contributors in privileged positions about the state of D\&I, particularly the challenges and bias on minority contributors. In the future, we plan to keep these longitudinal studies to continue monitoring and investigating the state of D\&I in the community, attracting more contributors' attention and eliminating challenges from D\&I-related bias.



\appendices
\bibliographystyle{IEEEtran}
\bibliography{bib}

%








\end{document}